# Non-Linear Modes in Lithium Niobate Ferroelectrics with niobium antisite defects

Atrayee Bhattacharya[1], Kamal Choudhary[2], Santanu Das[1,3], Loc Vu-Quoc[4], and

A. K. Bandyopadhyay[5]

1) Department of Metallurgy, BHU-IIT, Varanasi, India.

2) Dept of Materials Science, University of Florida, Gainesville, FL 32611, USA.

3) Department of Materials Science, University of North Texas, Denton, TX 76203, USA.

4) Department of Mechanical & Aerospacial Engineering, University of Florida, Gainesville, FL 32611, USA.

5) GCECT, W. B. University of Technology, Kolkata-700010, India
   **asisbanerjee1000@yahoo.co.in**

**Abstract**

Ferroelectric belongs to an important class of materials showing very interesting non-linear optical properties that have a variety of application in photonic devices In a discrete Hamiltonian, if we consider the space and time dependence of polarization vectors with an interaction term between two polarization domains, it gives rise to a nonlinear Klein-Gordon (K-G) equation that has been shown in some systems **[1]**. This equation as a governing equation enables multiple time-scale analysis (MTSA) to be performed on a real ferroelectric material that reveals new type of information on both linear (known as soft modes) and the nonlinear parts of frequencies and amplitudes of oscillations. MTSA was described in a K-G system based on staggered polarization where more emphasis was given to reveal the presence of intrinsic localized modes (ILM). Here, detailed results for the first three linear modes and their frequencies as a function of a wide range of niobium antisite defects or, impurities in lithium niobate ferroelectric are presented. It is found that when an external excitation (i.e. the input) is only harmonic in nature, it could not be ascertained whether the even modes of the solitonic waves (i.e. the output) will have phase-locking with this type of harmonic external force.

# I. Introduction

Ferroelectricity is an important property in solids that arises in certain crystals in terms of spontaneous polarization below Curie temperature **[2]**. Phenomenological models of ferroelectrics have been developed for computational engineering and applications **[3-5]**. An interesting characteristic is polarization reversal (switching) by an external driving force, as evident from the hysteresis curve of the polarization ($P$) and electrics field ($E$) vectors. This type of switching is important for non-volatile memory applications. Let us now talk about lithium niobate that have in recent years been the focus of intensive research because of their large optical nonlinearity and corresponding potential applications for devices such as a) four-phase mixing doublers, b) optical modulators, c) optical parametric oscillators, d) nonlinear frequency converters, e) second harmonic generators, f) holography, g) surface acoustic wave (SAW) devices, h) piezoelectric transducers, i) pyroelectric detectors, etc. **[1-4]**.

Periodically poled lithium niobate (PPLN) is one of the most versatile nonlinear optical materials over many years. However, PPLN is prone to thickness limitations: the risk of dielectric breakdown at higher poling field makes it unviable to obtain samples thicker than 0.05 cm. In the present literature search, it is found that the electric field for domain switching in near-stoichiometric lithium niobate (NSLN) ($\approx$ 50 mol% $Li_2O$) drops drastically as compared to that generally used congruent $LiNbO_3$ (CLN) ($\approx$ 48.50 mol% $Li_2O$). It was suggested that the non-stoichiometry, i.e. antisite defect or impurity, generally present as niobium antisite defects in such crystals, leads to large switching field **[1,6]**. Therefore, the present work was undertaken by also taking these impurities into consideration.

From the physics standpoint, ferroelectric behavior is commonly explained by the rotation of domains and domain walls that are present in the crystal with uniform polarization **[6-11]** (see references therein). To study the behavior of these domains and domain walls in lithium-niobate type of uniaxial ferroelectrics, a classical Hamiltonian was developed by taking the time variation through the kinetic part and the Landau-Ginzburg (L-G) polynomial in a potential formulation: (i) first by ignoring the depolarizing field and domain wall energy in the case of discrete polarization domains **[7]**, and (ii) second by considering these energy terms in the continuum limit **[8]**. In both cases, an interaction term ($k$) between the neighboring domains that is attached with the spatial part was considered. This effort gave rise to a nonlinear Klein-Gordon (K-G) equation. The elastic part was renormalized in the L-G polynomial in terms of expansion coefficients ($\alpha_1$ and $\alpha_2$).

These results were well known in the classical field theory. The K-G equation gives rise to soliton solutions like many other second-order differential equations, such as the KdV equation, the sine-Gordon equation, nonlinear Schrodinger equation (NLSE), etc.; these well-known equations are not discussed here. The soliton solutions based on the K-G equation showed features of shape and soliton velocity in lithium niobates **[8,9]** as well as in other nonlinear optical materials, such as metamaterials **[12,13]**. A very good account on soliton motion is given by Dauxois and Peyrard as well as by Kivshar and Agarwal **[14]**.

In the present work, we consider the 'domains' of length $L_1$ as rectangular boxes with uniform polarization in the *z*-direction (i.e. $P_z$) that are stacked sideways in the *x*-direction as per the general convention. A thin line separating two domains is the domain wall of width ($W_L$) **[15]**. Here, the polarization $P_i$ in the *i*th domain position interacts with $P_{i+1}$ in the (*i*+1)th domain and $P_{i-1}$ in the (*i*-1)th domain. This nearest-neighbor interaction term (denoted as *k*) plays a vital role in the overall oscillatory behavior as shown later. Hence, the stability of these oscillations also needs to be studied.

We note that the microscopic direction, along which ions or molecules in a crystal are re-arranged (during a structural phase transition) is generally described in terms of a collective vibrational mode of the lattice. In many cases, this mode is called a "soft" phonon mode due to its characteristically low frequency near the 'phase transition temperature'. It should be pointed out that the work on soft modes in various ferroelectrics dates back to 3-4 decades ago **[16-19]**. These investigations were directed either towards the understanding of the 'phase transition' and/or towards 'lattice instability' **[18-24]**.

It is not clear from the literature how the domains that are responsible for ferroelectric behavior give rise to nonlinear vibration modes. A detailed work via a perturbation technique is thus necessary to reveal the nonlinear behavior of different modes in terms of their amplitudes against frequencies. Moreover, before we analyse nonlinear modes, let us first study linear soft modes, whose frequency and the mode number should be related to some material parameters (e.g. coercive field, impurity content, etc.) in order to understand the underlying physics in the system. Therefore, the focus of the present work is to perform a detailed MTSA involving both time and spatial coordinates to relate the behavior of the extended modes to that of domains and domain walls which are important in ferroelectrics. The results would help to understand their switching behavior. As the K-G equation involves the variation of polarization (*P*) in both time (*t*) and space (*x*) variables at the second order, this equation could be ideal for carrying out such a study.

The paper is organized as follows: In **Section II**, a brief description of the MTSA technique is given, in both linear and nonlinear forms, and at the end of this section three cases are presented for electric field and damping. In **Section III**, the results and discussion are given for linear modes and frequencies that are shown to depend on impurity content of lithium niobate as well as on upper limit of domain wall width. In **Section III**, the nonlinear modes, their frequencies and amplitudes are also shown. In **Section IV**, the conclusions are given.

## II. Theoretical development:

**2.1. Perturbation technique:**

For the dynamics of polarization *P(x,t)*, the K-G equation with a damping term and driving ac electric field can be written as**:**

$$\frac{\partial^2 P}{\partial t^2} - \bar{k}\frac{\partial^2 P}{\partial x^2} + \bar{\gamma}\frac{\partial P}{\partial t} - \bar{\alpha}(P - P^3) - E_0 \cos(\Omega t) = 0 \qquad (1)$$

Various terms in Eq. (1) are explained in Appendix-I. Here, the driving field is dimensionless (i.e. $E/E_c$), based on the discrete Hamiltonian constructed for this purpose [7], where $E_c$ depends on the concentration of niobium antisite defects (see Ref. [15]) and there is a relation between the coercive or switching field and mole% impurity or defect density in lithium niobate ferroelectrics **[25,26]**. Hence, our Hamiltonian and consequently the K-G equation are impurity dependent, i.e., the present analysis is based on consideration of impurities normally present in such ferroelectrics (see later in the **Section III** for more details).

As in the L-G functional, the usual practice in this type of analysis is to account for only the odd-power terms in the polarization expansion **[8,15]**. Thus, let us consider the following expansion**:**

$$P = \varepsilon P_1 + \varepsilon^3 P_3 + \varepsilon^5 P_5 + \ldots \tag{2}$$

In addition, following multiple time-scale analysis **[15]**, we consider the time scales below as:

$$t_0 = t,\ t_2 = \varepsilon^2 t,\ \ldots,\ t_{2n} = \varepsilon^{2n} t \tag{3}$$

where the small parameter $\varepsilon = W_L/L_1$, where $L_1$ is the width of a domain and $W_L$ is the "domain wall" width. Note that $W_L$ can be of the order of few nanometers, and the sample may be considered as sort of nanowire with thickness $d$ ($d \ll L_2$, where $L_2$ is the sample length of the order of a few mm). With this definition of $\varepsilon$, we can study properties of the domains by means of MTSA, considering the time differential operators defined as follows**:** $D_0 = \partial/\partial t_0$, $D_2 = \partial/\partial t_2$, $D_4 = \partial/\partial t_4$, etc. By using the operators, we can write:

$$(D_0^2 + 2\varepsilon^2 D_2 D_0 + \ldots)(\varepsilon P_1 + \varepsilon^3 P_3 + \ldots) - \bar{k}(\varepsilon \frac{\partial^2 P_1}{\partial x^2} + \varepsilon^3 \frac{\partial^2 P_3}{\partial x^2} + \ldots) + \hat{\gamma}\varepsilon^2 (D_0 + \varepsilon^2 D_2 + \ldots)(\varepsilon P_1 + \varepsilon^3 P_3 + \ldots)$$
$$-\bar{\alpha}(\varepsilon P_1 + \varepsilon^3 P_3 + \ldots) + \bar{\alpha}(\varepsilon P_1 + \varepsilon^3 P_3 + \ldots)^3 - \hat{E}_0 \varepsilon^3 \cos(\Omega t_0) = 0$$
$$\tag{4}$$

Here, we define $\bar{\gamma} = \hat{\gamma}\varepsilon^2$ and $E_0 = \hat{E}_0\varepsilon^3$. Let us also take $\Omega = \omega + \varepsilon^2\Omega_0$. Then, Eq. (4) can be expressed as:

$$\cos(\omega t_0 + \varepsilon^2\Omega_0 t_0) = \cos(\omega t_0 + \Omega_0 t_2) = \frac{1}{2}\left(e^{i(\omega t_0 + \Omega_0 t_2)} + e^{-i(\omega t_0 + \Omega_0 t_2)}\right) \quad (5)$$

**2.2. Linear modes**

From Eq. (4), we obtain at the $\varepsilon^1$–order the equation as

$$D_0^2 P_1 - \bar{k}\frac{\partial^2 P_1}{\partial x^2} - \bar{\alpha} P_1 = 0 \quad (6)$$

Hence, following the derivation presented in Appendix 1, the frequency $\omega_L$ of the linear normal mode of vibrations can be written as

$$\omega_L^2 = n^2\pi^2 \bar{k} - \bar{\alpha} \quad (7)$$

**2.3. Nonlinear modes:**

From Eq. (4) we obtain the $\varepsilon^3$–order equation as

$$D_0^2 P_3 + 2D_2 D_0 P_1 - \bar{k}\frac{\partial^2 P_3}{\partial x^2} + \hat{\gamma} D_0 P_1 - \bar{\alpha} P_3 + \bar{\alpha} P_1^3 - \hat{E}_0 \cos(\omega t_0 + \Omega_0 t_2) = 0 \quad (8)$$

As shown in Appendix 2, we then arrive at the following relation:

$$(\omega I_2)\left(2\frac{d\hat{\sigma}}{dt} + \bar{\gamma}\hat{\sigma}\right) + (I_1 E_0)\sin\theta = 0 \quad (9)$$

Having made the above derivations, we now examine various specific cases to explain some elements of physics of both linear and nonlinear vibration modes in ferroelectrics.

**Case I:** $E_0 = 0$ and $\bar{\gamma} = 0$

Here, we are dealing with a case when there is no external excitation ($E_0 = 0$) and no damping ($\bar{\gamma} = 0$). Although being idealized, this situation may provide some insight into the theoretical aspects of the nonlinear analysis. From Eq. (9), taking $\frac{d\hat{\sigma}}{dt} = 0$ and oscillation amplitude $\hat{\sigma}$ as independent of time $t$, we can write the nonlinear part of the frequency $\theta$ as

$$\theta = \left(\frac{3\bar{\alpha}}{8\omega}\right)\left(\frac{I_4}{I_2}\right)(\hat{\sigma})^2 . t \tag{10}$$

Therefore, the frequency containing the nonlinear correction is:

$$\omega_{NL} = \omega_L + \left(\frac{3\bar{\alpha}}{8\omega_L}\right)\left(\frac{I_4}{I_2}\right)(\hat{\sigma})^2 \tag{11}$$

where $\omega_L$ is the linear frequency as given in Eq. (7), and $\hat{\sigma}$ is the oscillation amplitude that increases with $\omega_{NL}$, as shown below in Fig. 5 for the first three natural modes for near-stoichiometric lithium niobate. The integrals defined in Eq. (A9) are evaluated in Appendix I [Eqs. (A14), (A16) and (A17)].

**Case II : $E_0 = 0$ and $\bar{\gamma} \neq 0$**

Here, we are dealing with a case when there is no external excitation ($E_0 = 0$), but the system has a finite damping ($\bar{\gamma} \neq 0$). From Eq.(9), we obtain:

$$2\frac{d\hat{\sigma}}{dt} = -\bar{\gamma}\hat{\sigma} \tag{12}$$

Hence, the amplitude of oscillations decays in time as $\hat{\sigma} = e^{-(\bar{\gamma}/2)t}$. Again, from Eq.(9), we can write: $(2\omega I_2)\frac{d\theta}{dt} = \left(\frac{3\bar{\alpha}}{4}I_4\right)(\hat{\sigma})^2$ and $\frac{d\theta}{dt} = \left(\frac{3\bar{\alpha}}{8\omega}\right)\left(\frac{I_4}{I_2}\right)e^{-\bar{\gamma}t}$. Therefore, the solution can be written as:

$$\theta = -\left(\frac{3\bar{\alpha}}{8\omega\bar{\gamma}}\right)\left(\frac{I_4}{I_2}\right)e^{-\bar{\gamma}t} \quad (13)$$

Here, it is again seen that the 'nonlinear part' of the frequency tends to zero with the application of damping at large values of time *t*. The frequency of oscillations including the nonlinear term can now be written as:

$$\omega_{NL} = \omega_L + \theta = \omega_L - \left(\frac{3\bar{\alpha}}{8\omega_L \bar{\gamma}}\right)\left(\frac{I_4}{I_2}\right)\hat{\sigma}^2 \quad (14)$$

where $\omega_L$ is the linear frequency given in Eq. (7). Here, it should be noted that, in contrast to Case I, the amplitude of oscillations ($\hat{\sigma}$) will gradually tend to zero as $\omega_{NL}$ increases, as shown later in Fig. 6 for the first three natural modes for near-stoichiometric lithium niobate.

**Case III : $E_0 \neq 0$ and $\bar{\gamma} \neq 0$**

Let us now deal with a case when there is an external excitation ($E_0 \neq 0$), and the system has a finite damping ($\bar{\gamma} \neq 0$). As per the solution of Eq. (12), we can write the rate of change of the nonlinear part of the frequency and of the amplitude with respect to time as follows:

$$\frac{d\theta}{dt} = \left(\frac{3\bar{\alpha}}{8\omega}\right)\left(\frac{I_4}{I_2}\right)\hat{\sigma}^2 - \left(\frac{E_0}{2\omega}\right)\left(\frac{I_1}{I_2}\right)\frac{1}{\hat{\sigma}}\cos\theta \quad (15)$$

$$\frac{d\hat{\sigma}}{dt} = -\left(\frac{\bar{\gamma}}{2}\right)\hat{\sigma} - \left(\frac{E_0}{2\omega}\right)\left(\frac{I_1}{I_2}\right)\sin\theta \quad (16)$$

where $\omega^2 = \omega_L^2 = n^2\pi^2\bar{k} - \bar{\alpha}$, $I_1 = -\left(\dfrac{1}{n\pi}\right)[(-1)^n - 1]$, $I_2 = 1/2$, $I_4 = 3/8$, and $n \geq \dfrac{1}{\pi}\sqrt{\dfrac{\bar{\alpha}}{\bar{k}}}$.

These equations would be used in our analysis in the next section.

## III. Results and discussion

Here, an attempt was made to explore a perturbation technique in the case of actual ferroelectric materials, with the help of K-G equation involving the nonlinear L-G potential, for revealing the nature of the nonlinear modes with the contribution of the linear modes. Moreover, the effect of impurities or defects on the linear modes needs to be explored to produce results that may be applied in switching devices. As mentioned in **Section 2**, in the potential formulation of our Hamiltonian **[8,15]**, dimensionless $E$ contains switching field $E_c$ that again depends on the mole% impurity in ferroelectrics **[25,26]**. Our Hamiltonian thus takes impurities into consideration. Gopalan and coworkers **[27-29]** did an extensive work on the effect of impurities and domain wall motion in ferroelectrics, as was also done by Shur et al. **[30],** who dealt with lithium antisite vacancy.

In the present study, the electric field is oscillatory ($E = E_0 \cos(\omega t)$) that varies with time. The "polarization domains" formed are 'perturbed' by this external oscillatory field ($E$), and it will lead to 'rearrangements' of the domains. Thus, the single 'soft mode' vibration will be transferred to multiple 'soft mode' vibrations. Note that we are working much below the Curie point at a fixed temperature (say, at room temperature). So, the creation of multiple modes of vibrations does not require an additional lattice degree of freedom. Thus, these modes are "automatically" created by the perturbation with the external oscillatory field. First of all, let us consider the linear modes in lithium niobate.

**3.1. Analysis of the linear modes:**

In Ref. **[25]**, Yan et al. showed that the coercive field in lithium niobate increases with its impurity content (expressed in mole%) possibly due to stiffening of the domains, and thus the internal resistance to oscillations (i.e., damping) also increases with the impurity **[11]**. The coercive field and the impurity data are taken from Ref. **[25]** and **[26]**, which are used in the present analysis.

It is quite interesting to use some experimental data of a ferroelectric having applications in nonlinear optical devices. For lithium niobate crystal, known experimental values are $E_c$ = 40 kV/cm and $P_s$ = 0.75 C/m², and we calculate the value of $\bar{\alpha}$ = 353.42 **[6,8]**. If we assume that the value of the interaction term is $\bar{k}$ = 1, then from Eq. (7) we obtain:

$$\omega_L^2 = n^2\pi^2 - 353.42 \quad (\omega_L \geq 0) \tag{17}$$

The above derivation of Eq. (17) yields a value of $n \geq 6$, since $\omega_L^2 \geq 0$. The natural frequency $\omega_L$ of each of the first three linear oscillation modes with the above data and based on Eq.(7) is given below: For $n=6$, $\omega_L = 1.42$ (the first-mode frequency), for $n=7$, $\omega_L = 11.41$ (the second-mode frequency), and for $n=8$, $\omega_L = 16.68$ (the third-mode frequency).

In this manner, different experimental values of the coercive field $E_c$ can be taken from Yan et al. **[25]** and Tian et al. **[26]** for lithium niobate and lithium tantalate respectively, and the values of $\bar{\alpha}$ can be evaluated to estimate the frequency of different linear modes of oscillations, using Eq. (7). These frequencies represent the transverse optical (TO) phonon modes, i.e., so-called "soft" modes. These soft modes have a lower frequency due to the 'interplay' between the local restoring force and the dipolar interaction; both of which have been considered in our Hamiltonian.

At each impurity content in mole%, there is a corresponding value of poling field ($E_c$) that in turn determines Landau coefficient ($\alpha$) at each mode number ($n$) corresponding to the frequency. Hence, for the three natural modes there are three sets of data. The variation of the first three normal modes is plotted in Fig. 2 (the best-fitted lines are drawn through + (higher mode), squares (the next lower mode) and then circles (the lowest natural mode)) against a wide range of impurity content, which is quantified by the coercive field ($E_c$) only for lithium niobate, but not for lithium tantalate since the corresponding plot is very similar to that of lithium niobate. To note that the mode number sharply decreases with the impurity content or $E_c$ up to a value that is equal to 40 kV/cm, at which Gopalan et al. **[6]** found it easier to work with the hysteresis study with the appropriate thickness of the sample. After this value, the curve asymptotically decreases towards higher impurity content (i.e., the congruent side).

This sharp fall of mode number at low impurity content might indicate the start of the pinning of the domains and from where the rotation of the domains could start becoming difficult and then the coercive field $E_c$ starts increasing. From the analysis of the linear modes of vibrations (or, the so-called 'soft' modes), this result might be treated as a confirmation of domain rotations getting somewhat stiffer, and thus requiring a "higher energy" for such rotations from this particular value of $E_c$, albeit there are no data on the mechanical deformation strain to back up this inference at the moment. The linear frequency is plotted against impurity content in Fig.3 (the best-fitted lines are drawn through various points with upper curve denoting the higher-mode number). The scatter in the data points are less evident in Fig. 3 compared to that in Fig. 2. It is quite clear from Fig. 3 that the linear frequency ($\omega_L$) shows a minimum, i.e. minimum energy at $E_c$ = 40 kV/cm (i.e. corresponding to an impurity content = 0.133 mole% of niobium antisite vacancy), thereby indicating a stable situation, which is relatively more prominent in case of the first natural mode of oscillations (see the lower curve in Fig.3). Gopalan et al **[6,28]** found this particular sample easier for switching that also shows 'lowest energy' as evident from Fig.3. This may be considered as a confirmation of a general behavior of a large number of experimental work done on this material. However, from one of the most significant parameters (i.e. the phonon hopping coefficient) derived from the two-phonon bound state in LN ferroelectric against impurity content also showed a sharp transition at this point that confirms our approach to the understanding of the pinning behavior through linear vibrations **[31]**.

As mentioned in Section 1, the domain walls are important to understand ferroelectric phenomena. By taking the static soliton solution of Lines and Glass **[32]**, we obtained a critical zone of stability for dimensionless polarization $P_c$ between 1 and $1/\sqrt{3}$ after linearization through a Jacobian transformation, and found two limits of half the domain wall width (DWW). At $P_c$ = 1, the lower DWW was almost constant (the most stable situation) at 0.75 nm, i.e., 1.5 times the lattice spacing of lithium niobate which agrees well with that of Padilla et al **[33]**, and at $P_c$ = $1/\sqrt{3}$, the upper limit of DWW (written as 'upper wall width' in Fig. 4) was 210 nm that corresponds well with that of Gopalan et al. **[6]** (see Eq. (26a) in Ref. **[7]**). This upper limit could be considered as the 'thermodynamic limit' that corresponds to the theoretical limit of the coercive field in the $2^{nd}$ order phase transitions, which was explained long time ago by Jona and Shirane **[34]**. As DWW is inversely proportional to the poling field, an effort is made here to derive an approximate relation with it.

From different $E_c$ values at different impurity contents, we can work out this upper limit. The plot of the mode number against this DWW for the first three natural modes in the linear regime is shown in Fig. 4. The upper curve represents higher mode numbers. It is seen that all the modes at the initial stage increase rather steeply and then slowly move towards the stoichiometric side. Such property of the linear natural modes of (domain) vibrations is significant and a new information for domain wall width. This analysis is approximate, as we have assumed a static soliton in Ref. **[7]** by taking only the spatial variation of the domain-wall width. Nevertheless, it shows a trend in our data for linear vibrations vis-a-vis poling field that could be important for applications in various devices. Note that Phillpot and coworkers **[35]** used density functional theory (DFT) approach and molecular dynamics simulations to investigate the behavior of Er defects in lithium niobate that actually showed the role of defects on the charge balance having some relevance to our work on impurity dependence of linear modes, although DFT calculation in our system is beyond the scope of the present work.

However, our recent work **[1]** on discrete energy levels of the bright soliton that is estimated through a hypergeometric function showed that they have a certain number of energy levels that depend upon the internal obstacle, i.e., damping in the system. There is no obstacle in a region that is characterized as 'energy gap' where the dipoles can be rotated by forward and reverse poling. The corresponding domain walls are relatively steeper for forward poling than that of reverse poling, thus establishing the existence of energy levels of solitons **[28,36]**. This might have some consequences for second harmonic generation in lithium niobate ferroelectrics **[37]**. Next, we analyze the nonlinear modes in the system.

**3.2. Analysis of the nonlinear modes**

First of all, it should be emphasized here that the study of nonlinear vibrations has significance due to its applications for describing classical and quantum discrete breathers in the K-G lattice **[31,38,39]** for targeted energy transfer and quantum computation. It has also many other applications in different fields of technology, namely, a) 'phase-coherent optical pulse synthesis' **[40]**, b) 'parametric light generation' **[41]**, c) 'ultrafast spectroscopy' **[42]**, etc. Concerning the latter application, extensive investigations by Nelson and coworkers **[43]** on both lithium tantalate and lithium niobate need a particular mention. Some of these applications have relevance to switching in ferroelectric materials that depends on the poling or switching field. Thus, this switching field was related to the linear vibrations in the previous section.

For a lithium niobate crystal, from Eq. (11), the amplitude of oscillations is plotted against the nonlinear frequency in Fig. 5 for Case-I (i.e., no external excitation and no damping), where the amplitude is seen to increase for a sample with $E_c = 40$ kV/cm. The first three natural modes are only shown here. From Eq. (14), the amplitude of oscillations is plotted against the nonlinear frequency in Fig. 6 for Case-II (i.e., no external excitation, but with finite damping), where the amplitude is seen to decrease steadily with the nonlinear frequency, as expected, since there is an effect of damping. From the exponential relation written after Eq. (12), it is seen that the amplitude of oscillations tends to zero with the application of damping at large time ($t$).

For Case-III, from Eqs. (15) and (16), at $E_c$ = 40 kV/cm, it is necessary to note that the electric field $E_0$ does not affect the 1$^{st}$ and 3$^{rd}$ (i.e., $n$ = odd) modes of nonlinear oscillations, whereas the situation is reverse in case of 2$^{nd}$ and 4$^{th}$ (i.e., $n$ = even) modes. Hence, when $n$ is odd, the situation is the same as described in Case II, i.e., the amplitude and nonlinear frequency can be described by Eqs. (13) and (14), respectively. It implies that the solitons will be guided differently by two different sets of equations for even and odd modes respectively.

The above property of the solitons cannot be properly explained from an important work of Salerno and Zolotaryuk **[44]** on the effect of a biharmonic a.c. driver on the solitonic motion, albeit that work was based on the sine-Gordon equation. According to that work, the solitons will have motion under moderate to high damping if the a.c. driver itself has odd modes, but for the even modes, there will be 'phase-locking' between this external oscillatory force and the solitons. As in the Case III ($E_0 \neq 0$ and $\bar{\gamma} \neq 0$), however, the external excitation (i.e., the input) is only harmonic in nature, and it cannot be ascertained whether the even modes of the solitonic waves (i.e.. the output) will have phase-locking with this type of harmonic external force. Nevertheless, these even modes will still have usefulness with their amplitude going relatively high with smaller time scale, as per the relation: $\hat{\sigma} = e^{-(\bar{\gamma}/2)t}$.

Fig. 7 is a plot of the amplitude against nonlinear frequency (Case III) at two different non-dimensional electric fields, $E_0$ = 1 and $E_0$ = 100, with a moderate value of the damping term $\bar{\gamma} = 0.5$, and keeping the non-dimensional time at $t'$ = 20 in Eqs. (15) and (16). In both these cases, the amplitude increases with frequency, except that at higher field value, there are still some oscillations (see the upper curve). These curves are drawn for the odd mode at $n = 7$ for the same sample at $E_c = 40$ kV/cm. To note that in the case of oscillations for odd modes (Fig.7), the amplitude increases steadily with the nonlinear frequency, but at higher field ($E_0 = 100$) the amplitude increases with some oscillations at the beginning and these oscillations disappear at higher value of nonlinear frequency.

It should be mentioned that in the temporal dynamics in lithium niobate, with the progressive application of a static field, the Landau potential starts loosing its symmetry, as it was evident from the phase plane diagram at higher external excitation **[10]**. Case III gives an indication that the motion for the soliton is quite probable in lithium niobate for odd modes, as shown in Fig.7. It is pertinent to mention that for an oscillatory field, the Lyapunov exponent spectrum became positive at non-dimensional field value of $E_0 = 130$ indicating chaos in lithium niobate system, which was also evident from the phase plane diagram at this value **[10]**. More work need to be done in the future to find a relation between the suppression of chaos and different behavior of the even modes with the application of a harmonic oscillatory field in the context of multiple time-scale analysis.

## IV. Conclusions

The multiple time-scale analysis gives us a proper insight into the linear and non-linear modes of oscillations, under different conditions of external driving force and damping, which is based on a Klein-Gordon equation with the Landau potential. The first three natural modes in the linear regime show an important variation with the impurity content indicating the start of some kind of stiffening of the rotation of domains and domain walls at $E_c = 40$ kV/cm. These mode numbers show an increasing trend with the upper limit of half the domain wall width that is worked out approximately from our previous perturbation model, indicating the effect of the domain-wall width on the linear oscillations in the system. Equivalently, it shows the functional dependence on the poling field that has consequence for switching applications in ferroelecrics.

In the nonlinear regime, without any field and damping, the amplitude of oscillations increases with the nonlinear frequency, while it decreases when some damping is applied without any external excitation. For the latter case, the amplitude becomes significant at very low time, when the soliton motion will prevail. When both the field and damping are applied in lithium niobate, only the 'odd' modes of nonlinear frequency of oscillations will be present, while those with the 'even' modes will have different type of behavior and usefulness. This could be due to phase-locking between the harmonic external excitation and these 'even' modes, and this might be related to some kind of progressive breaking of the symmetry of the Landau potential with field that was previously observed in our dynamic system analysis **[10]**.

**Acknowledgements**

The authors (AKB and SD) would like to thank Professor Venkat Gopalan of Dept. of Materials Science and Engineering at Penn State University (USA) for providing some experimental data and useful discussion during the early part of this work. The authors would also like to specially thank Professor Boris Malomed of Faculty of Engineering of Tel Aviv University (Israel) for many corrections in the manuscript and for very helpful suggestions during preparation of this paper.

**<u>Appendix - I</u>**

Eq.(1) in the main text contains all the non-dimensional terms as: $P = P'/P_s$, where $P_s$ is the saturation polarization in C/m² (typical value for LiNbO$_3$ ferroelectrics as 0.75 C/m² [6]), the normalized field is $E_0 = E'/E_c$, where $E_c$ is the coercive field (when $P = 0$) in kV/cm in the usual nonlinear hysteresis curve of $P$ vs. $E$ with a typical value for the same material as 40 kV/cm [6], $t = t'/t_c$, where $t_c$ is considered as the critical time scale for polarization to reach a saturation value, i.e. at or near the domain walls that are of importance to our study with a typical value of 10 ns for a switching time of (say) 200 ns for a damping value $\bar{\gamma} = 0.50$ [10] that are based on the above data, and $x = x'/W_L$, where $W_L$ = domain wall width of the order of a few nanometers. Eq. (1) is obtained after dropping the prime notation, and by taking $\alpha_1 = \alpha_2/P_s^2$ and $\bar{\alpha}_1 = \bar{\alpha}_2 = \bar{\alpha} = (\alpha_{12} P_s)/E_c$ [6,15]. Here, the critical time is defined as:

$$t_c = \frac{1}{Q_d}\sqrt{\frac{m_d P_s}{E_c}}$$

seconds, where $Q_d$ and $m_d$ are the inertial constants [15]. Moreover, the interaction between the domains and damping term that gives resistance to oscillations in the case of discrete polarization [8] are expressed as: $\bar{k} = \frac{kP_s}{2E_c}$ and $\bar{\gamma} = \frac{\gamma P_s}{t_c E_c}$, where $\gamma$ is the damping coefficient which is a decay constant relating to the loss of polarization due to internal friction during its motion in a ferroelectric and the rest of the terms are defined above.

Here, let us seek a solution of Eq.(6) of the form :

$$P_1 = [A(t_2)e^{i\omega t_0} + c.c.]X(x) \tag{A1}$$

where *c.c* stands for the complex conjugate of $A(t_2)e^{i\omega t_0} = \bar{A}(t_2)e^{-i\omega t_0}$. From Eq.(6), we obtain:

$$\bar{k}\frac{d^2X}{dx^2} = -(\omega^2 + \bar{\alpha})X \tag{A2}$$

The solution of Eq.(A2) with the boundary condition $P_1(t, 0) = 0$ at $x = 0$ implies that $X(0) = 0$, which yields:

$$X(x) = B_1 \sin\left(\sqrt{\frac{\omega^2 + \bar{\alpha}}{\bar{k}}}\,.x\right) \tag{A3}$$

And we obtain the following dispersion relation as:

$$\omega^2 + \bar{\alpha} = n^2\pi^2\bar{k} \tag{A4}$$

where n = 1, 2, 3, ..., i.e., positive integers.

**Appendix II**

Now, let us seek a solution of Eq.(8) of the form :

$$P_3 = [A(t_2)e^{i\omega t_0} + c.c.]Y(x) \tag{A5}$$

From Eq.(8), by taking $A'(t_2) = \dfrac{dA}{dt_2}$ and by equating the coefficients of $e^{i\omega t_0}$ to zero, we obtain:

$$\bar{k}A\frac{d^2Y}{dx^2} + (\omega^2 + \bar{\alpha})AY(x) = 2(i\omega)A'X(x) + \hat{\gamma}(i\omega)AX(x) + \bar{\alpha}(3A^2\bar{A})[X(x)]^3$$
$$-\frac{\hat{E}_0}{2}[\cos(\Omega_0 t_2) + i\sin(\Omega_0 t_2)] \tag{A6}$$

By multiplying both sides of Eq.(A6) by $X(x)$ and integrating from 0 to 1, we obtain the resulting left-hand side (LHS) of this equation as follows:

$$\text{LHS} = A[\int_0^1 Y\left(\bar{k}\frac{d^2X}{dx^2} + (\omega^2 + \bar{\alpha})X\right)dx] = 0 \qquad (A7)$$

As a result of an integration by parts and the boundary conditions $Y(0) = Y(1) = 0$, by an application of Eq.(A2), the resulting right-hand side (RHS) of Eq.(A6) becomes

$$2(i\omega)A'\int_0^1 X^2 dx + \hat{\gamma}(i\omega)A\int_0^1 X^2 dx + \bar{\alpha}(3A^2\bar{A})\int_0^1 X^4 dx - \frac{\hat{E}_0}{2}(\cos\Omega_0 t_2 + i\sin\Omega_0 t_2)\int_0^1 X dx = 0$$

$$(A8)$$

Let us denote the above integrals as follows:

$$I_1 = \int_0^1 X dx, \quad I_2 = \int_0^1 X^2 dx, \quad \text{and} \quad I_4 = \int_0^1 X^4 dx \qquad (A9)$$

Then, Eq.(A8) takes the following form:

$$(i\omega)I_2[2A' + \hat{\gamma}A] + (3\bar{\alpha})(A^2\bar{A})I_4 - \frac{\hat{E}_0}{2}(\cos\Omega_0 t_2 + i\sin\Omega_0 t_2)I_1 = 0 \qquad (A10)$$

By taking the oscillation amplitude σ(t₂) and the non-linear frequency θ(t₂) as real quantities, and also by defining the factor $A(t_2)$ in Eq.(A5), we get:

$$(i\omega)I_2[2\sigma' + 2i\theta'\sigma + \hat{\gamma}\sigma] + \left(\frac{3}{4}\bar{\alpha}\right)\sigma^3 I_4 - \hat{E}_0 I_1(\cos\theta - i\sin\theta)[\cos(\Omega_0 t_2) + i\sin(\Omega_0 t_2)] = 0$$

$$(A11)$$

By equating the imaginary and the real parts, each separately to zero, we obtain:

$$(2\omega\sigma I_2)\theta' = \left(\frac{3}{4}\bar{\alpha}I_4\right)\sigma^3 - (\hat{E}_0 I_1)[\cos(\theta - \Omega_0 t_2)] \qquad (A12)$$

and,

$$\omega I_2(2\sigma' + \hat{\gamma}\sigma) + \hat{E}_0 I_1[\sin\theta\cos(\Omega_0 t_2) - \cos\theta\sin(\Omega_0 t_2)] = 0 \qquad (A13)$$

Since $t_2 = \varepsilon^2 t$ by virtue of Eq.(3), and if we define $\hat{\sigma} = \varepsilon\sigma$, let us also take $\Omega_0 = 0$ and by letting $E_0 = \varepsilon^3 \hat{E}_0$ and $\bar{\gamma} = \varepsilon^2 \hat{\gamma}$, we then derive Eq.(9).

The integrals defined in Eq.(A9) can now be evaluated as follows. First, using Eq.(A4), we obtain $I_1$:

$$I_1 = \int_0^1 X dx = \int_0^1 \sin\left(\sqrt{\frac{\omega^2 + \bar{\alpha}}{\bar{k}}}.x\right) dx = -\left(\frac{1}{n\pi}\right)[(-1)^n - 1] \tag{A14}$$

Since $\omega^2 \geq 0$, Eq.(A4) leads to the mode number $n$ being expressed as:

$$n \geq \frac{1}{\pi}\sqrt{\frac{\bar{\alpha}}{\bar{k}}} \tag{A15}$$

Next, the second integral $I_2$ is evaluated as**:**

$$I_2 = \int_0^1 X^2 dx = \int_0^1 \sin^2\left(\sqrt{\frac{\omega^2 + \bar{\alpha}}{\bar{k}}}.x\right) dx = \frac{1}{2} \tag{A16}$$

Finally, the third integral $I_4$ can be evaluated as**:**

$$I_4 = \int_0^1 X^4 dx = \int_0^1 \sin^4\left(\sqrt{\frac{\omega^2 + \bar{\alpha}}{\bar{k}}}.x\right) dx = \frac{3}{8} \tag{A17}$$

### **Figure Captions**

**Fig. 1 :** Schematic diagram of the polarization domains in the discrete case (as per Ref. [14]).

**Fig. 2 :** The mode number (*n*) for the first three normal modes of linear vibrations vs. the impurity content (mole%) in lithium niobate [the best fitted curves are drawn through computed points: + (higher mode number), squares (next lower mode number) and then circles (the lowest normal mode number)].

**Fig. 3 :** The linear frequency ($\omega_L$) for the first three normal modes against impurity content (mole%) in lithium niobate. [best fitted lines are drawn through various computed points with upper curve denoting the higher mode number].

**Fig. 4 :** The mode number for the first three normal modes of linear vibrations vs. the upper limit of half the domain wall width (nm) in lithium niobate. The upper curve shows this behavior for higher mode number.

**Fig. 5 :** Amplitude against non-linear frequency [Eq.(45)] in lithium niobate (Case I). Blue continuous line ($\omega = 1.41$, $n = 6$), ------ dashed line ($\omega = 11.41$, $n = 7$), and dashed dotted line --.--.-- ($\omega = 16.68$, $n = 8$) for $\bar{k} = 1$, $\bar{\alpha} = 3.5294 \times 10^2$, $E_0 = 0$, and $\bar{\gamma} = 0.0$.

**Fig. 6 :** Amplitude against non-linear frequency (Eq.(55)) in lithium niobate (Case II) Blue continuous line ($\omega = 1.41$, $n = 6$), ------ dashed line ($\omega = 11.41$, $n = 7$), and dashed dotted line --.--.-- ($\omega = 16.68$, $n = 8$) for $\bar{k} = 1$, $\bar{\alpha} = 3.5294 \times 10^2$, $E_0 = 0$, and $\bar{\gamma} = 0.5$.

**Fig. 7 :** Amplitude against non-linear frequency (Eq.(55)) in lithium niobate (Case III) $E_0 = 1$, $n = 7$, $\omega = 11.41$ for (----- dashed line) and $E_0 = 100$, $n = 7$, $\omega = 11.41$ (blue continuous line) for $\bar{k} = 1$, $\bar{\alpha} = 3.5294 \times 10^2$ and $\bar{\gamma} = 0.5$ (for $n = 6$ and $n = 8$, see Case II). Initial conditions: At time t=0, θ = -π/2 and $\hat{\sigma}$ =0.00001 [for solving equations (56) & (57)].

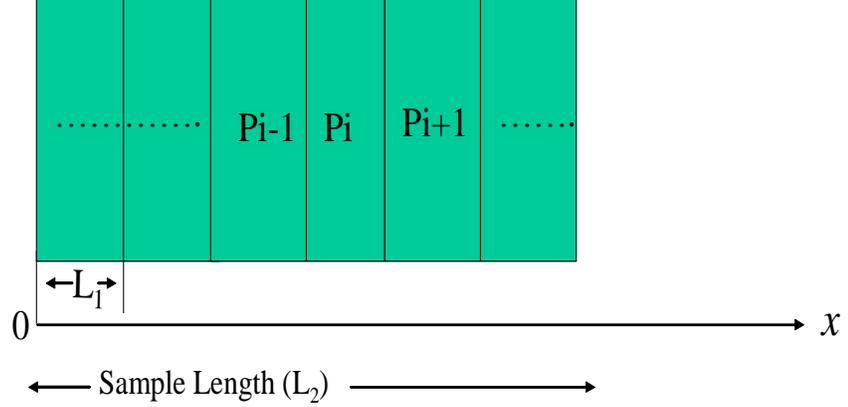

**Fig.1 :** Schematic diagram of the polarization domains in the discrete case (as per Ref.[14])

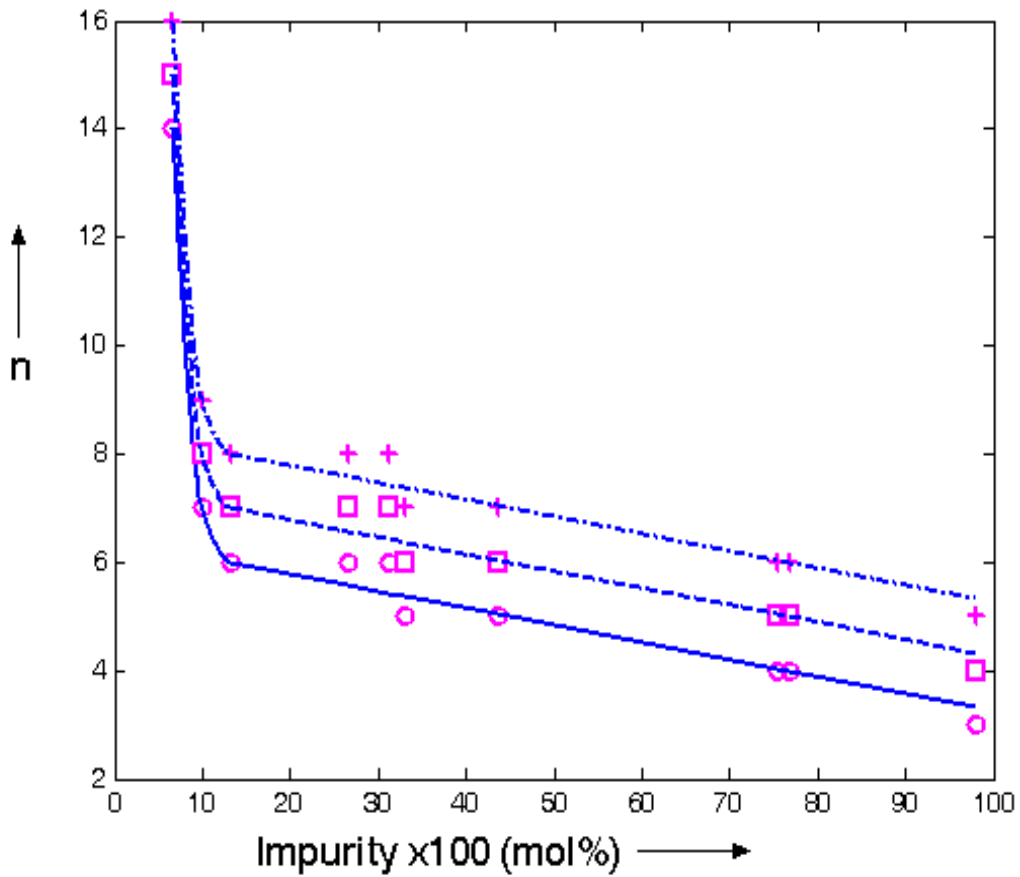

**Fig. 2 :** The mode number (*n*) for the first three normal modes of linear vibrations vs. the impurity content (mole%) in lithium niobate [the best fitted curves are drawn through computed points: + (higher mode number), squares (next lower mode number) and then circles (n=6)].

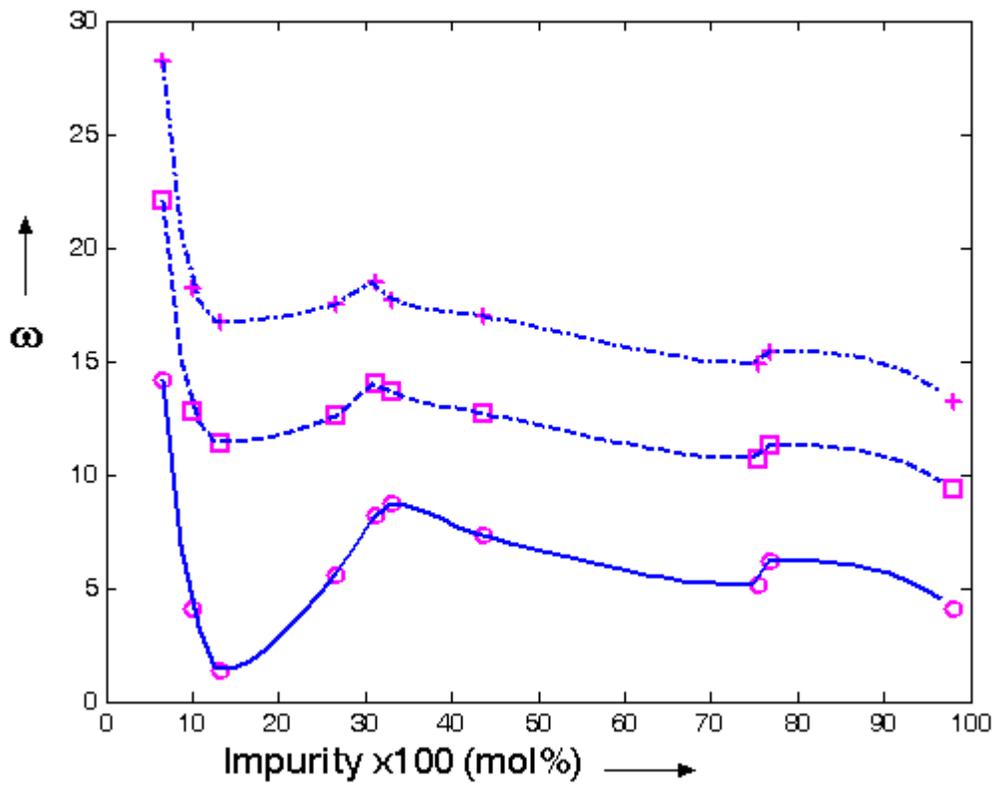

**Fig. 3 :** The linear frequency ($\omega_L$) for the first three normal modes against impurity content (mole%) in lithium niobate. [the best fitted lines are drawn through various computed points with upper curve denoting the higher mode number].

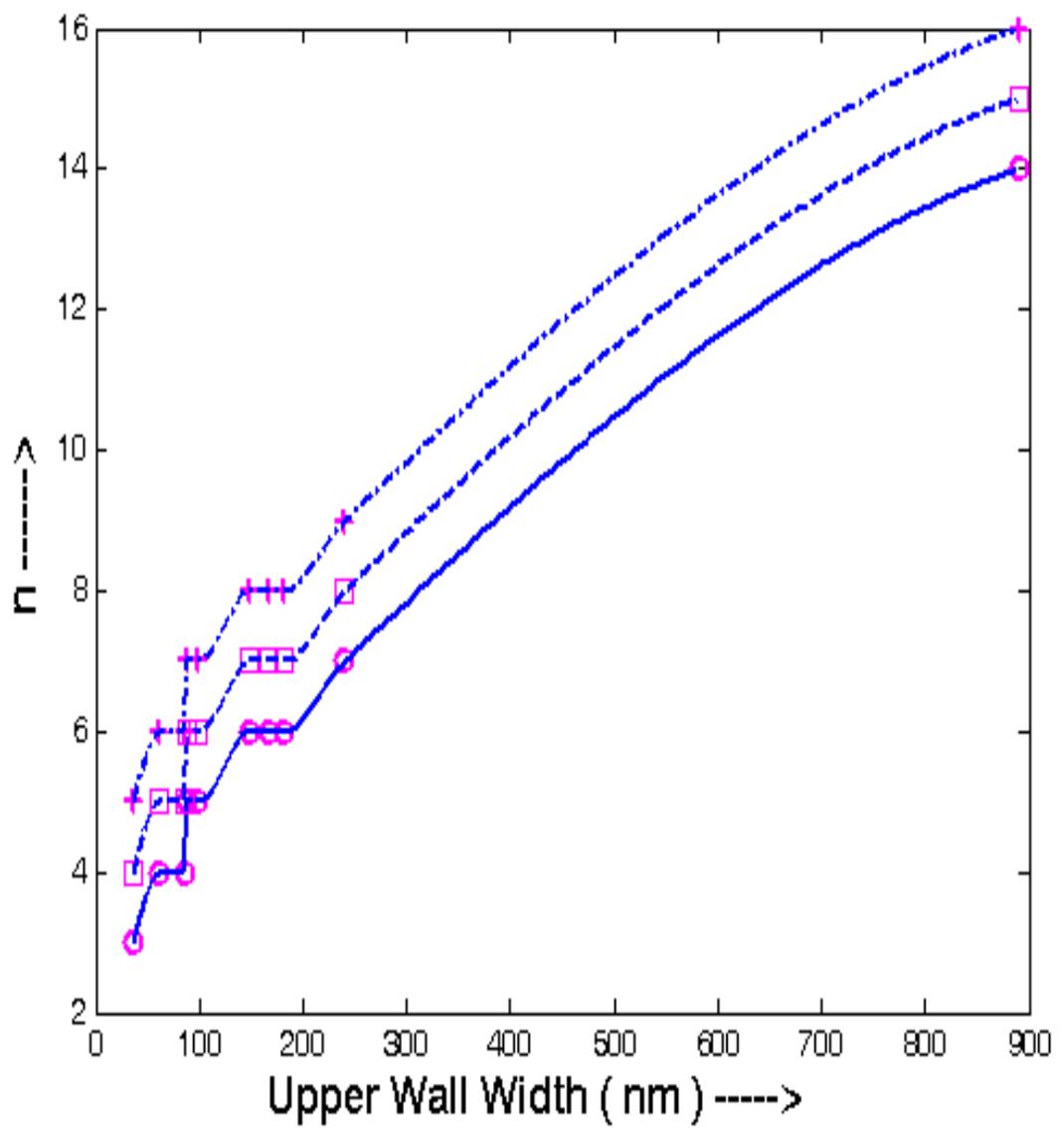

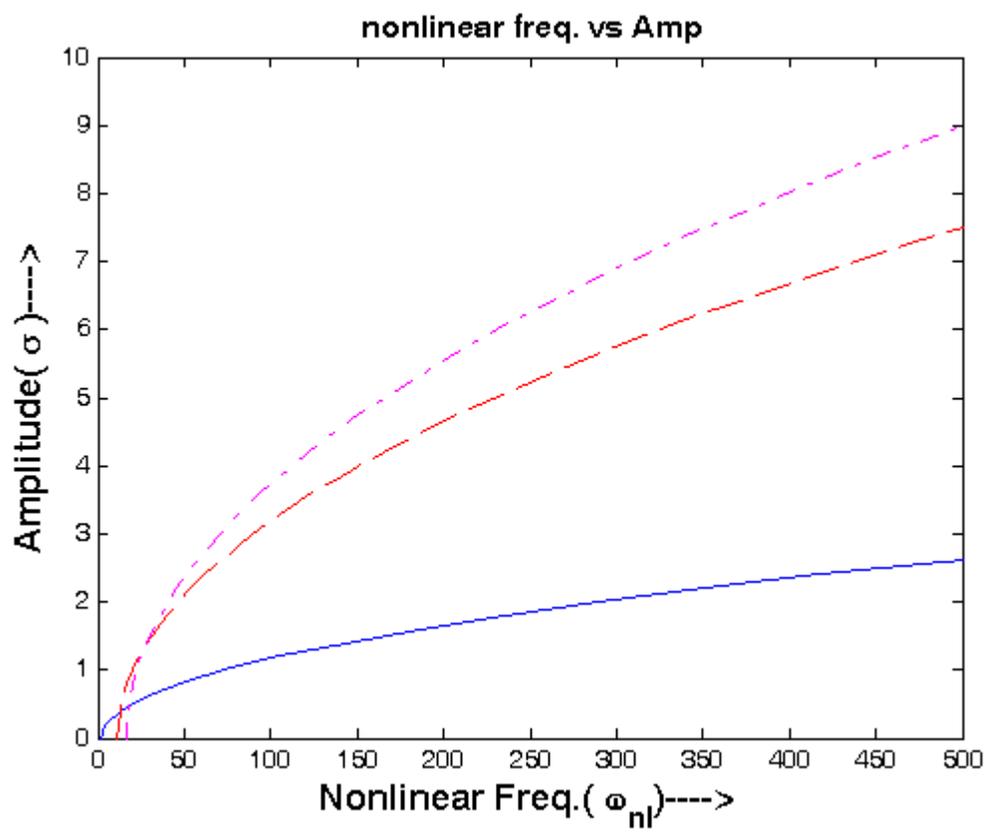

**Fig. 5 :** Amplitude against non-linear frequency [Eq.(11)] in lithium niobate (Case I). Blue continuous line ($\omega = 1.41$, $n = 6$), ------ dashed line ($\omega = 11.41$, $n = 7$), and dashed dotted line --.--.-- ($\omega = 16.68$, $n = 8$) for $\bar{k} = 1$, $\bar{\alpha} = 3.5294 \times 10^2$, $E_0 = 0$, and $\bar{\gamma} = 0.0$.

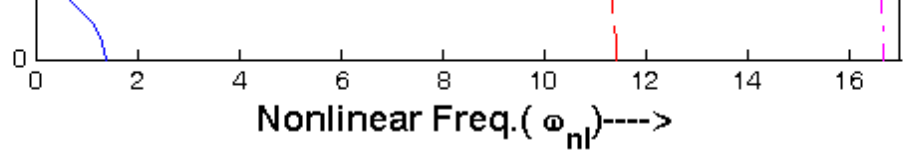

**Fig. 6 :** Amplitude against non-linear frequency (Eq.(14)) in lithium niobate (Case II) Blue continuous line ($\omega = 1.41$, $n = 6$), ------ dashed line ($\omega = 11.41$, $n = 7$), and dashed dotted line --.--.-- ($\omega = 16.68$, $n = 8$) for $\bar{k} = 1$, $\bar{\alpha} = 3.5294 \times 10^2$, $E_0 = 0$, and $\bar{\gamma} = 0.5$.

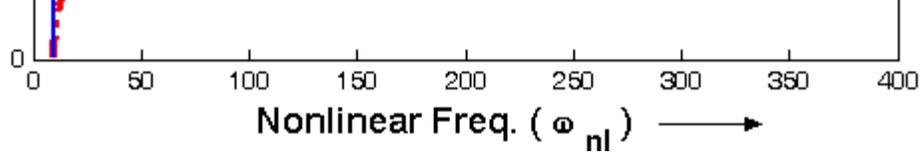

**Fig. 7 :** Amplitude against non-linear frequency [Eq.(14)] in lithium niobate (Case III) $E_0 = 1, n = 7, \omega = 11.41$ for (----- dashed line) and $E_0 = 100, n = 7, \omega = 11.41$ (blue continuous line) for $\bar{k} = 1, \bar{\alpha} = 3.5294 \times 10^2$ and $\bar{\gamma} = 0.5$ (for $n = 6$ and $n = 8$, see Case II). Initial conditions: At time t=0, $\theta = -\pi/2$ and $\hat{\sigma} = 0.00001$ [for solving equations (15) & (16)].